\pdfoutput=1
\documentclass[preprint,12pt]{elsarticle}
\usepackage{amsmath}
\usepackage{bm}
\usepackage{amssymb}
\usepackage[ruled]{algorithm2e} % For algorithms
\usepackage{algorithmic}
\usepackage{color}% This package is for using color package.
\newcommand{\red}[1]{\textcolor{red}{#1}}% Chang-Dong uses red text to state the revision or revision note.
\newcommand{\blue}[1]{\textcolor{blue}{#1}}% The others use blue text to state their revision or revision note.

\usepackage{framed}
\usepackage{lipsum}

\journal{Knowledge-Based Systems}

\begin{document}

\begin{frontmatter}



\title{PRSI: Privacy-Preserving Recommendation Model Based on Vector Splitting and Interactive Protocols\tnoteref{label1}}\tnotetext[label1]{\blue{The work was supported by National Key Research and Development Program of China (2021YFF1201200) and NSFC (62276277).}}


\author{Xiaokai~Cao\fnref{label2}}
\ead{caoxk@mail2.sysu.edu.cn}

\author{Wenjin~Mo\fnref{label2}}
\ead{mowj27@mail2.sysu.edu.cn}

\author{Zhenyu~He\fnref{label2}}
\ead{hezhy65@mail2.sysu.edu.cn}

\author{Changdong~Wang\corref{cor1}\fnref{label2}}
\ead{changdongwang@hotmail.com}

\cortext[cor1]{Corresponding author.}

\address[label2]{School of computer Science and Engineering Sun Yat-sen University, Guangzhou, china.\fnref{label1}}

%\affiliation{organization={},%Department and Organization

\begin{abstract}
With the development of the internet, recommending interesting products to users has become a highly valuable research topic for businesses. Recommendation systems play a crucial role in addressing this issue. To prevent the leakage of each user's (client's) private data, Federated Recommendation Systems (FedRec) have been proposed and widely used. However, extensive research has shown that FedRec suffers from security issues such as data privacy leakage, and it is challenging to train effective models with FedRec when each client only holds interaction information for a single user. To address these two problems, this paper proposes a new privacy-preserving recommendation system (PRSI), which includes a preprocessing module and two main phases. The preprocessing module employs split vectors and fake interaction items to protect clients' interaction information and recommendation results. The two main phases are: (1) the collection of interaction information and (2) the sending of recommendation results. In the interaction information collection phase, each client uses the preprocessing module and random communication methods (according to the designed interactive protocol) to protect their ID information and IP addresses. In the recommendation results sending phase, the central server uses the preprocessing module and triplets to distribute recommendation results to each client under secure conditions, following the designed interactive protocol. Finally, we conducted multiple sets of experiments to verify the security, accuracy, and communication cost of the proposed method.
\end{abstract}




\end{frontmatter}



\section{Introduction}\label{sec:introdution}

With the continuous development of society, more and more products and services are available for people to choose. Therefore, it is increasingly difficult for people to find what they are really interested in, which is the problem of information overload. To address the above issues, recommendation systems (RS) have been proposed and widely used~\cite{DBLP:journals/internet/SmithL17, DBLP:journals/tmis/Gomez-UribeH16}. Researchers have developed various kinds of recommendation systems to effectively uncover users' latent interests~\cite{DBLP:journals/air/DauS20,DBLP:conf/recsys/DavidsonLLNVGGHLLS10}.

However, RS typically require users' historical interaction data, to recommend items (such as products, movies, or short videos) of interest to users. The interaction data may contain sensitive personal information, and its direct use could potentially compromise user privacy. With the advancement of data science and increasing attention from regulations (e.g., GDPR) and citizens regarding the collection and use of personal data, balancing the efficient use of data with privacy security has become a significant research issue.

Federated Learning is a key technology for addressing this issue~\cite{mcmahan_communication-efficient_2017}. Training recommendation systems using a federated learning framework allows for a balance between model performance and security to some extent. This is also the mainstream research direction in privacy-preserving recommendation systems, known as Federated Recommendation Systems (FedRec). Perifanis et al.~\cite{DBLP:journals/kbs/PerifanisE22} and Lin et al.~\cite{DBLP:journals/jpdc/LinLDQPR23} designed different FedRec frameworks based on neural network collaborative filtering algorithms. Han et al.~\cite{DBLP:conf/www/Han0ML21}, Zhou et al.~\cite{DBLP:journals/tkde/ZhouWGGZ21}, and Imran et al.~\cite{DBLP:journals/tois/ImranY0NZ023} extended FedRec to models suitable for time-series data, which is a common data type in real-world scenarios.

Although FedRec avoids directly transmitting data and achieves distributed training by transferring gradient parameters, numerous studies have shown that this approach is still not secure. Researchers have found that by attacking the gradient parameters, it is possible to recover users' private data~\cite{zhu_deep_2019, yang_using_2023, DBLP:conf/infocom/WangSZSWQ19}. Moreover, methods such as poisoning attacks can even manipulate recommendation results without relying on prior knowledge~\cite{DBLP:conf/sigir/YuanNHCY23, DBLP:conf/ijcai/RongHC22}.

Although some researchers have employed techniques such as differential privacy, homomorphic encryption, and generating fake interactions to enhance data security in FedRec~\cite{DBLP:conf/recsys/MintoHLH21, DBLP:journals/expert/ChaiWCY21, DBLP:journals/expert/LinLPM21}, the security issues in FedRec remain a topic of concern. Moreover, another challenge lies in distributed training when each client only holds the interaction data of a single user in FedRec. Due to the highly sparse nature of the interaction data matrix, if each client only holds the interaction data of a single user, the data on each client will exhibit strong heterogeneity. This significantly increases the difficulty of training and reduces the model's performance. Additionally, since each client only has access to one user's interaction data, the amount of local data is extremely small, making it difficult to train an effective model with such limited data. However, this issue is critical in practical applications of recommendation systems, where each client often only possesses their own personal interaction data. Therefore, it is crucial to address the recommendation problem in which each client holds the interaction data of only a single user.

To address this issue, we propose a privacy-preserving recommendation model based on vector splitting and interactive protocols (PRSI). This approach consists of a preprocessing module and two main phases: the collection of interaction data by the central server (i.e. clients send interaction information) and the distribution of recommendation results by the central server (i.e. clients receive recommendations). Specifically, we have made the following contributions:
\begin{itemize}
    \item We designed a preprocessing module called vector splitting, which splits the users' interaction vectors and the recommendation vectors into multiple random vectors, and generates fake interaction items to protect the real interaction data and recommendation results.

    \item In the interaction data collection phase, we designed an interactive protocol to securely upload user interaction data while ensuring privacy. This protocol first utilizes the preprocessing module to protect the interaction data. Then, multiple clients engage in interaction following a predefined random communication method, which protects their ID information and IP addresses. Finally, the interaction data is aggregated at the central server.
    \item In the recommendation result distribution phase, we designed an interactive protocol to securely transmit the recommendation results. This protocol uses the triplets generated in the previous phase to reversely track the clients' IP addresses and employs a semi-random communication method to protect the privacy of the clients' information.
\end{itemize}

Compared to existing FedRec, this method offers the following advantages:
\begin{itemize}
    \item Privacy Protection: It utilizes fake interaction data and secure multi-party computation protocols to safeguard user privacy effectively.

    \item Realistic Applicability: It is suitable for scenarios where each client holds interaction data for only a single user, addressing a common real-world situation.

    \item Flexibility: It is not limited to specific types of recommendation models and can be seamlessly integrated with commonly used recommendation models without significant adjustments.
\end{itemize}

The other sections of this paper are organized as follows: Section~\ref{sec: Related Work} reviews recent research on the security aspects of FedRec. Section~\ref{sec: Proposed Method} provides a detailed description of the method proposed in this paper. Section~\ref{sec: Experiments} outlines the experimental design used to validate the effectiveness and security of the proposed method. Section~\ref{sec: Conclusion} summarizes the research findings and discusses potential future research directions.

%At present, the main process of recommendation can be summarized as follows: collecting the interaction information of all users; training the recommendation model based on the above interaction information; recommending a certain number of items for each user. In this mode, each user needs to share the historical interaction data with a central server, so that the central server can train the model. In general, the interaction data is sent directly by the user to the central server. However, there is a privacy risk in the way of sending data directly to the central server. Due to this mode, the central server can obtain the correspondence between the user identity and the specific interaction information. Specifically, the server can learn about the interaction data for any user, which damage the privacy of users' interaction.

\section{Related Work}\label{sec: Related Work}

Model training in recommendation systems often involves the privacy data of multiple users, making it particularly suitable to use distributed training methods across multiple clients for training recommendation models. Consequently, federated learning has been widely applied to distributed training in recommendation systems in recent years. In 2020, Muhammad et al.~\cite{DBLP:conf/kdd/MuhammadWOTSHGL20} proposed a fast distributed learning model called FedFast, based on the classic federated learning framework~\cite{mcmahan_communication-efficient_2017}. This model selects a subset of clients for parameter aggregation in each training round and sends the aggregated model to other clients, thus enhancing the efficiency of training FedRec. In 2022, Perifanis et al.~\cite{DBLP:journals/kbs/PerifanisE22} introduced Federated Neural Collaborative Filtering (FedNCF), the first method to train neural network-based collaborative filtering algorithms within a federated learning framework. In 2023, Lin et al.~\cite{DBLP:journals/jpdc/LinLDQPR23} proposed a new federated recommendation algorithm, FedNeuMF, by enhancing the classic neural collaborative filtering algorithm. This method leverages auxiliary user information and item attributes to improve recommendation accuracy. Additionally, strategies such as matrix factorization\cite{DBLP:journals/kbs/WanZLFSG22, DBLP:journals/kbs/ZhangLCZ19}, neural networks~\cite{DBLP:conf/icml/MaiP23} and heterogeneous models~\cite{DBLP:conf/icde/YuanQCT0Y24} have also been applied to the research of FedRec.

In practical applications of recommendation systems, data often exist in the form of time series. Therefore, time series-based federated recommendation tasks have become another research hotspot in this field. Han et al.~\cite{DBLP:conf/www/Han0ML21} adapted time series recommendation tasks to the federated learning framework, proposing On-device Deep Learning for Privacy-Preserving Sequential Recommendation (DeepRec). Zhou et al.~\cite{DBLP:journals/tkde/ZhouWGGZ21} constructed an item clustering tree for handling time series data, built a social recommendation system, and applied it to train on large-scale social data. Imran et al.~\cite{DBLP:journals/tois/ImranY0NZ023} developed ReFRS, a FedRec based on variational autoencoders to learn users' temporal preferences.

However, extensive research indicates that the practice of transmitting gradient parameters in federated learning faces security issues such as gradient leakage, where user privacy information can be recovered through gradient parameter attacks~\cite{zhu_deep_2019, yang_using_2023, DBLP:conf/infocom/WangSZSWQ19}. As FedRec have been studied in greater depth, more scholars have designed corresponding attack methods. For example, to address the client heterogeneity issue in FedRec, a common approach is to cluster or group clients; however, studies have shown that this method can leak user privacy information~\cite{DBLP:conf/www/HeLKH24}. Significant research has been dedicated to attacks on FedRec. Yuan et al.~\cite{DBLP:conf/sigir/YuanNHCY23} proposed a poisoning attack that does not rely on any prior knowledge, using a set of malicious users to upload toxic gradients to manipulate the ranking and exposure rate of target items in top-K recommendations. Rong et al.~\cite{DBLP:conf/ijcai/RongHC22} designed poisoning gradients based on random approximation and hard user mining strategies, using these poisoned gradients to corrupt the global model. Poisoning attacks are widely used in FedRec due to their ability to manipulate recommendation results at a very low cost. Novel poisoning attacks have been proposed by Yin et al. and Zhang et al. in~\cite{DBLP:conf/www/YinXFG24} and~\cite{DBLP:journals/tkde/ZhangYY24}, respectively. In addition, inference attacks have also been applied to FedRec~\cite{DBLP:conf/www/YuanYNCHY23}.

To ensure the security of FedRec, scholars have proposed a series of protective measures, primarily incorporating two types of technologies. One type involves traditional encryption methods such as differential privacy and homomorphic encryption, while the other involves techniques specifically designed for recommendation systems, such as fake interaction methods. Minto et al.~\cite{DBLP:conf/recsys/MintoHLH21} proposed using differential privacy techniques to perturb data, thereby protecting user data security. Chai et al.~\cite{DBLP:journals/expert/ChaiWCY21} demonstrated that publicly available gradient parameters could be used to recover users' original rating information and applied homomorphic encryption to FedRec, performing matrix decomposition in a homomorphic encryption environment to safeguard privacy data. Wang et al.~\cite{DBLP:journals/vldb/WangYCYZZ22} introduced a meta-learning approach to extract feature information and then used differential privacy to protect the recommendation model, thus ensuring the data security of the participating clients. Lin et al.~\cite{DBLP:journals/expert/LinLPM21} proposed generating fake interaction items through user averaging and mixed filling methods, using these fake interactions to obfuscate the information attackers could acquire, thus enhancing federated recommendation security. Lin et al.~\cite{DBLP:conf/recsys/LinP021} used Fake Marks and Secret Sharing techniques to modify the gradient parameters of the model to achieve privacy protection. Liang et al.~\cite{DBLP:conf/aaai/LiangP021} argued that fake interactions and ratings can interfere with model training and affect model performance, so they proposed a lossless FedRec that removes noise in a privacy-preserving manner while balancing security and model performance. Yuan et al.~\cite{DBLP:conf/icde/YuanYQNLY24} proposed a FedRec model that eliminates the need to transmit parameters to avoid privacy leakage through gradient parameters.

Although many researchers have focused on the security issues of FedRec, absolute data security cannot yet be guaranteed. In addition to security issues, training a FedRec becomes significantly more challenging when each client is assigned interaction data from only one user. However, the scenario where each client corresponds to a single user is precisely the issue faced in real-world applications. To address this problem, we propose a privacy-preserving recommender system based on vector splitting and interactive protocols. This method is designed for scenarios where each client holds interaction vectors for only one user and employs secure multi-party computation protocols to ensure the security of user privacy data.

\section{The Proposed Method}\label{sec: Proposed Method}

\subsection{Problem Statement}
There are $N$ clients and a central server involved in a recommender system task in privacy computing. In this problem, client $i$ holds only one interaction vector $\bm{u_i}$, and $\bm{u_i}$ contains the indices of the items that client $i$ has interacted with. The central server needs to collect all clients' interaction vectors $\bm{u_i}, i = 1,2,..., N$ to form an interaction matrix, then use the interaction matrix to train the recommendation model for recommending the corresponding items for each client, and finally send the recommendation result back to each client. In this task, there are the following requirements:

\begin{itemize}
    \item The interaction vector for each client must be sent accurately to the central server so that the central server can train the recommendation model.

    \item In order to protect the client's privacy, the client's IP information must be protected so that the central server cannot trace the client corresponding to the interaction vector $\bm{u_i}$. However, the recommendation results need to be sent to the corresponding client.

    \item Each client cannot obtain the IP information, interaction vector $\bm{u_i}$ and recommendation results of other clients.
\end{itemize}

\subsection{Overview of the PRSI framework}

In this paper, we design a privacy-preserving recommendation model based on vector splitting and interactive protocols (PRSI). This method addresses the distributed training problem where each client only possesses a single interaction vector. We consider privacy protection from two aspects. In one aspect, the interaction data, recommendation results, and ID information of the clients must not be disclosed to other clients. To achieve this, we design a preprocessing module to protect these data. In another aspect, while the interaction data of the clients can be accessed by the central server to ensure the accuracy of the recommendation results, the ID information of the clients must remain confidential to the central server. This protects the privacy of the clients by safeguarding their ID information.

This method includes a preprocessing module (vector splitting) and two main stages (interactive protocols). The function of the preprocessing module is to use fake interaction information to protect the real interaction information (or recommendation information) and split the processed interaction vectors or recommendation vectors (i.e., the recommendation results for each user) into multiple random vectors. This module is detailed in section \ref{subsec: Splitting Interaction Vector}.

The first main stage is the central server collecting the clients' interaction vectors (that is, the clients upload their interaction vectors), which is detailed in section \ref{subsec: Collecting Interaction Vectors}. In this stage, all clients use the preprocessing module to split their interaction vectors into random vectors. Then, following the process outlined in the interactive protocol \ref{Alg:Collecting Interaction Vectors}, they send/receive random vectors. This multiple-round random communication hides and protects the clients' ID information and IP addresses. Finally, the random vectors are sent to the central server, which trains the recommendation model.

The second main stage involves the central server sending the recommendation results to each client (that is, the clients download the recommendation results), as detailed in section \ref{subsec: Sending Recommendation Results}. In this stage, the central server uses the preprocessing module to split the recommendation item vectors into random vectors. Then, following the process outlined in the interactive protocol \ref{Alg:Sending Recommendation Results}, these random vectors are sent to the clients, delivering the recommendation results to them.

\subsection{Vector Splitting}\label{subsec: Splitting Interaction Vector}

\begin{figure*}[!t]%
\centering
\includegraphics[width=1.0\textwidth]{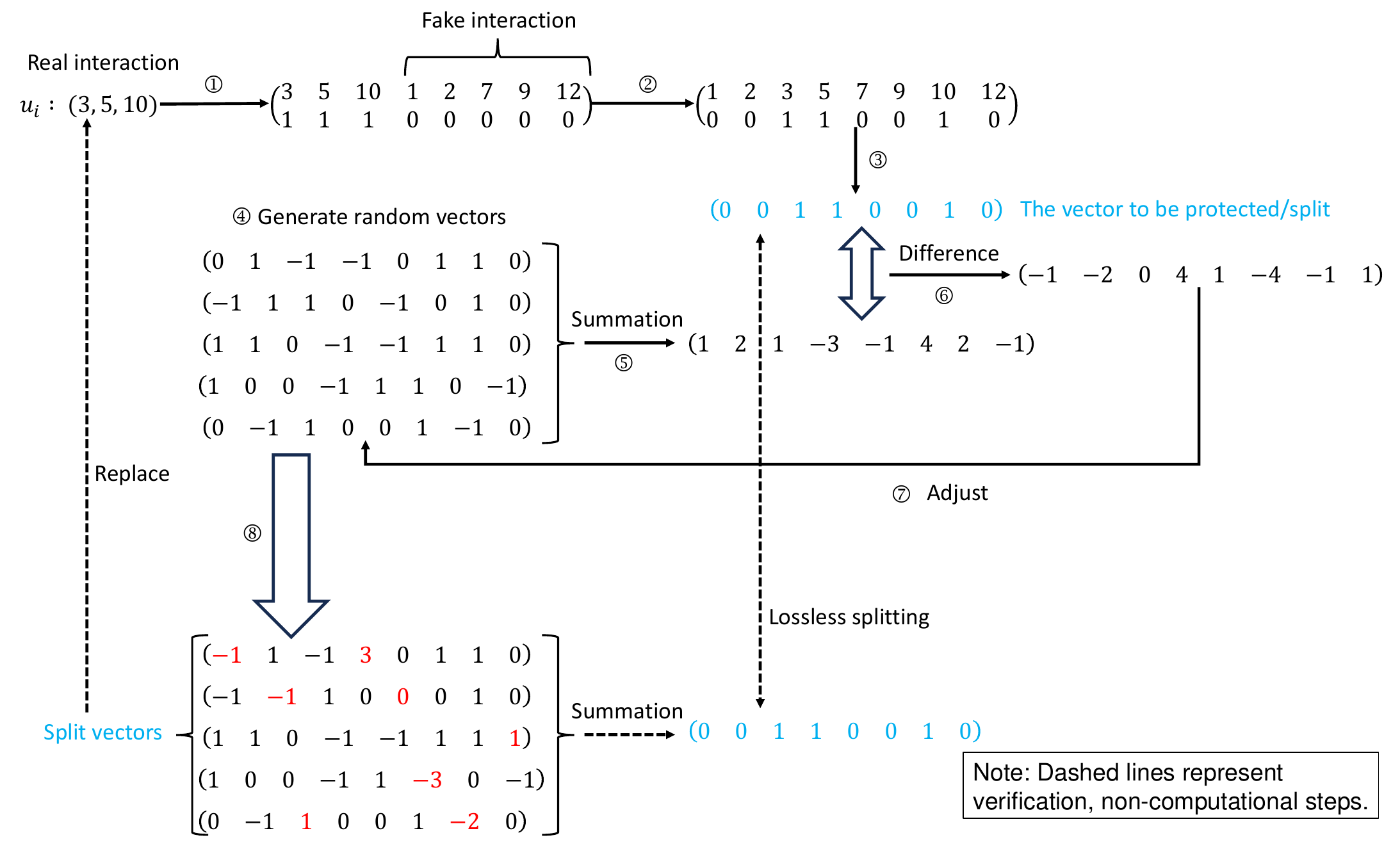}\\
\caption{Basic steps of Vector Splitting.}\label{Fig-VectorSplitting}
\end{figure*}

In this section, we introduce the preprocessing module, namely the concept and method of 'Vector Splitting'. The purpose of splitting vectors is to protect the local interaction vectors of clients (i.e., users) and the recommendation results generated by the central server. Without loss of generality, we illustrate this process basing on interaction vectors. Figure~\ref{Fig-VectorSplitting} illustrates the basic steps of this process.

Let the total number of interaction items be $N_{item}$, the maximum number of interaction items for each user be $n_{max}$, each interaction vector is split into $S_{spl}$ vectors.

To protect the interaction data from being leaked, we randomly select some non-interacted (fake interaction) items to mask the real interaction items. For an interaction vector $\bm{u_i}$, let $n^* = c\cdot n_{max}$, where $c\in N^+, c>1$, and satisfy $ c\cdot n_{max} < N_{item}$, and $n^* - len(\bm{u}_i)$ represents the number of fake interaction items. The function $len(\cdot)$ calculates the dimensionality of the corresponding vector. This process corresponds to Step $1$ in \figurename~\ref{Fig-VectorSplitting}.

Then, the indices of fake interaction items and real interaction items are shuffled, and the result is denoted as matrix $\mathbf{u}_i^*$, i.e.,
\begin{eqnarray}\label{u_i^*}
\mathbf{u}_i^*=shuffle
        \left(
        \left[\begin{array}{cc}
            \bm{u}_i & randint(1, N_{item}; n^* - len(\bm{u}_i)) \\
            ones(len(\bm{u}_i)) & zeros(n^* - len(\bm{u}_i))
        \end{array}\right]
        \right),
\end{eqnarray}
where $\bm{u}_i$ represents the items interacted by user $i$. The function $randint(1, N_{item}; n^* - len(\bm{u}_i))$ generates a vector of dimension $n^* - len(\bm{u}_i)$ with elements being distinct integers from $1$ to $N_{item}$, ensuring they do not overlap with the elements in $\bm{u}_i$. Its purpose is to generate fake interaction items to mask the actual interaction items $\bm{u}_i$. The functions $ones(\cdot)$ and $zeros(\cdot)$ respectively generate vectors of all ones and all zeros corresponding to the given dimensionality. In this context, $1$ is used to mark the real interaction items in the matrix's first row, while $0$ is used to mark the fake interaction items. The function $shuffle(\cdot)$ denotes the random permutation of elements in the first row of the corresponding matrix, with the second row undergoing corresponding permutations aligned with the first row. This process corresponds to Step $2$ in \figurename~\ref{Fig-VectorSplitting}.

The first row of the matrix $\mathbf{u}_i^*$ contains the indices of the real and fake interaction items, denoted as vector $\bm{u}_{i,1}^*$. The second row contains the boolean values indicating whether the items are real interaction items, denoted as $\bm{u}_{i,2}^*$. To protect the interaction information, we split $\bm{u}_{i,2}^*$. This process corresponds to Step $3$ in \figurename~\ref{Fig-VectorSplitting}.

First, randomly generate $S_{spl}$ splitting vectors $\bm{V}_i^s$ with values of $\{0, 1, -1\}$, i.e.,
\begin{eqnarray}\label{V_i^s}
\bm{V}_i^s = rand(0,1,-1; n^*),
\end{eqnarray}
where, the function $rand(0,1,-1; n^*)$ generates an $n^*-$dimensional vector with elements from the set $\{0,1,-1\}$. This process corresponds to Step $4$ in \figurename~\ref{Fig-VectorSplitting}.

Then, we calculate the difference between the sum of the splitting vectors $\sum _s \bm{V}_i^s$ and the interaction vector $\bm{u}_{i,2}^*$, i.e., let:
\begin{eqnarray}\label{D_diff}
\bm{D}_{diff} = \bm{u}_{i,2}^* - \sum _s \bm{V}_i^s.
\end{eqnarray}
This process corresponds to Step $5$ and $6$ in \figurename~\ref{Fig-VectorSplitting}. 
Then, randomly select $len(\bm{D}_{diff})$ splitting vectors $\bm{V}_i^s, s\in [1, 2, \ldots, S_{spl}]$, and use the components of the difference $\bm{D}_{diff}(d), d=1,2,\ldots, len(\bm{D}_{diff})$ to adjust these splitting vectors, i.e.,
\begin{eqnarray}\label{V_i^s d}
\bm{V}_i^s (d) = \bm{V}_i^s (d) + \bm{D}_{diff}(d).
\end{eqnarray}
The adjusted splitting vectors are the final splitting vectors, satisfying $\bm{u}_{i,2}^* = \sum _s \bm{V}_i^s$. The tuple $(\bm{V}_i^s, \bm{u}_{i,1}^*),s=1, \ldots, S_{spl}$ contains the interaction information of user $i$. This process corresponds to Step $7$ and $8$ in \figurename~\ref{Fig-VectorSplitting}. The basic process of the splitting vectors is shown in Algorithm \ref{Alg:Splitting interaction vector}.

\begin{algorithm}[!t]
	\caption{Vector Splitting}
	\label{Alg:Splitting interaction vector}
	\begin{flushleft}
		\textbf{Input: } Interaction vector or recommended vector $\bm{u}_i$, the number of items $N_{item}$, maximum number of items per user interaction $n_{max}$, the number of split vectors $S_{spl}$.
	\end{flushleft}
	\begin{algorithmic}[1]
		\STATE Let $n^* = c\cdot n_{max}$, where $c\in N^+, c>1$, and satisfy $ c\cdot n_{max} < N_{item}$.
        \STATE Generate the interference interaction information $\mathbf{u}_i^*$ by Eq.~\eqref{u_i^*}.
        %\STATE Denote the first row of the matrix $\mathbf{u}_i^*$ as $\bm{u}_{i,1}^*$ and the second row as $\bm{u}_{i,2}^*$.
        \FOR{ $s = 1$ to $S_{spl}$}
		\STATE $\bm{V}_i^s = rand(0,1,-1; n^*)$,
		\ENDFOR
        \STATE Let $\bm{D}_{diff} = \bm{u}_{i,2}^* - \sum _s \bm{V}_i^s$,
        \FOR{$d$ in $range(len(\bm{D}_{diff}))$}
        \STATE Choose a vector $\bm{V}_i^s$, let $\bm{V}_i^s (d) = \bm{V}_i^s (d) + \bm{D}_{diff}(d)$.
        \ENDFOR
	\end{algorithmic}
        \textbf{Output: } Split vectors tuples $(\bm{V}_i^s, \bm{u}_{i,1}^*),s=1, \ldots, S_{spl}$ of vector $\bm{u}_i$.
\end{algorithm}

\subsection{Collecting Interaction Vectors}\label{subsec: Collecting Interaction Vectors}

In the previous section, we split the interaction vectors into multiple splitting vectors. In this section, we will describe the process by which the central server collects these splitting vectors and explain the necessity of this process.

If the splitting vectors $\bm{V}_i^s$ are sent directly to the central server, the server could trace the IP addresses to identify which client the interaction information belongs to, which is an undesirable outcome. On the other hand, if clients hide their IP addresses, the central server cannot send the recommendation results back to the clients. Therefore, to balance security and feasibility, we designed an interactive algorithm to facilitate data collection.

In section \ref{subsec: Splitting Interaction Vector}, each client splits its local interaction vector into $S_{spl}$ tuples. Based on this, each client $i$ generates a virtual ID, denoted as $V_{ID}^i$, forming a triple $(V_{ID}^i, \bm{V}_i^s, \bm{u}_{i,1}^*)$. The virtual ID $V_{ID}^i$ is used to tag the tuples $(\bm{V}_i^s, \bm{u}_{i,1}^*)$ belonging to the same client $i$, facilitating the subsequent distribution of recommendation results.

The process of collecting interaction vectors involves three main operations: sending data, receiving data, and collecting data. During the data sending operation, each client sends the corresponding data according to specific rules. In the data receiving operation, each client records and saves the received data according to predefined protocols. In the data collection operation, the central server aggregates the split vectors and trains the recommendation model.

Firstly, during the data sending operation, each client determines the target of transmission probabilistically using an exponentially decaying probability denoted by $R_{sto}\in (0,1)$. Specifically, with a probability of $R_{sto}$, all triples $(V_{ID}^i, \bm{V}_i^s, \bm{u}_{i,1}^*)$ held locally are individually and randomly sent to other clients; otherwise, they are sent to the central server and the data sending phase terminates.

Then, during the data receiving operation, each client receives triples sent by other clients and retrieves their corresponding IP addresses $I_{IP}^j$.  If these IP addresses are not already saved locally, the tuple $(V_{ID}^j, I_{IP}^j)$ is stored in the local database; otherwise, no action is taken. This operation significantly enhances the efficiency of subsequent recommendation result transmissions, as will be elaborated in the following section.

\begin{algorithm}[!t]
	\caption{Collecting Interaction Vectors}
	\label{Alg:Collecting Interaction Vectors}
	\begin{flushleft}
		\textbf{Input: } Interaction vector $\bm{u}_i$ for client $i, i=1,2, \ldots, N_{user}$, where $N_{user}$ is the number of client/user, attenuation coefficient $\alpha$.
	\end{flushleft}
	\begin{algorithmic}[1]
		\STATE \texttt{Client $i$ do:}
        \STATE Generate the local IP address $I_{IP}^i$ and virtual ID $V_{ID}^i$, and set $p_{sto}=1$.
        \STATE Use Algorithm \ref{Alg:Splitting interaction vector} to split $\bm{u}_i$ and generate tuples $(\bm{V}_i^s, \bm{u}_{i,1}^*)$.
        %\definecolor{shadecolor}{rgb}{0.5,0.92,0.5}
%\begin{shaded}
        \WHILE{$1$}
        \STATE \texttt{Client $i$ do:}
        \STATE Generate a random number $R_{sto}\in (0,1)$,
        \IF{$P(R_{sto} < p_{sto})$}
            \STATE  send the triplet $(V_{ID}^i, \bm{V}_i^s, \bm{u}_{i,1}^*)$ to any arbitrary client,
        \ELSE
            \STATE send the triplet $(V_{ID}^i, \bm{V}_i^s, \bm{u}_{i,1}^*)$ to the central server,
        \ENDIF
        \STATE $p_{sto}=p_{sto}*\alpha$,
        \ENDWHILE
        %\end{shaded}
        %\definecolor{shadecolor}{rgb}{0.5,0.92,0.5}
%\begin{shaded}

        \WHILE{$1$}
        \STATE \texttt{Client $i$ do:}
        \STATE Receive the triplet $(V_{ID}^j, \bm{V}_j^s, \bm{u}_{j,1}^*)$, and obtain the IP $I_{IP}^j$ of client $j$,
        \IF{$I_{IP}^j \notin LD$}
            \STATE Save the tuples $(V_{ID}^j, I_{IP}^j)$ to the local database $LD$,
        \ENDIF
        \ENDWHILE
        %\end{shaded}
        %\definecolor{shadecolor}{rgb}{0.5,0.92,0.5}
%\begin{shaded}

        \STATE \texttt{Central server do:}
        \STATE Store the triples $(V_{ID}^i, \bm{V}_i^s, \bm{u}_{i,1}^*)$ in the local database $ID_{list}$.

        \FOR{$V_{ID}^i \in ID_{list}$}
        \STATE Use Eqs.~\eqref{u_i,2^**}-\eqref{tilde u_i^*} to calculate $\mathbf{u}_i^*$, and remove the redundant information to obtain $\tilde{\bm{u}}_i$.
        \ENDFOR
%\end{shaded}
        \STATE Train the recommendation model using $\tilde{\bm{u}}_i$ to obtain $\bm{Rec}_i$.
	\end{algorithmic}
        \textbf{Output: } The recommended items $\bm{Rec}_i$ for user $i, i=1,2, \ldots, N_{user}$.
\end{algorithm}

After multiple rounds of data interactions, as the probability value $R_{sto}$ gradually diminishes, all triples will eventually be sent to the central server. Finally, the central server aggregates the received triples $(V_{ID}^i, \bm{V}_i^s, \bm{u}_{i,1}^*)$ by accumulating the vectors $\bm{V}_i^s$ from triples with the same virtual ID $V_{ID}^i$, i.e.,
\begin{eqnarray}\label{u_i,2^**}
\bm{u}_{i,2}^{**} = \sum _s \bm{V}_i^s,
\end{eqnarray}
and let
\begin{eqnarray}\label{tilde u_i^*}
\tilde{\bm{u}}_i =
        \left[\begin{array}{c}
            \bm{u}_{i,1}^* \\
            \bm{u}_{i,2}^{**}
        \end{array}\right].
\end{eqnarray}
Finally, remove the redundant information to obtain $\tilde{\bm{u}}_i$, that is, remove the columns containing $0$ from the matrix $\mathbf{u}_i^*$, then delete the second row of the matrix. Denote the resulting matrix as $\tilde{\bm{u}}_i$. Thus, $\tilde{\bm{u}}_i$ represents the interaction vector of client/user $i$, equivalent to $\bm{u}_i$.

Central server aggregate all $\bm{u}_i$ vectors to form the user-item interaction matrix, and using the interaction matrix to train the recommendation model, the recommendation results are denoted as $\bm{Rec}_i$. The basic procedure for collecting interaction vectors is illustrated in Algorithm \ref{Alg:Collecting Interaction Vectors}.

\subsection{Sending Recommendation Results}\label{subsec: Sending Recommendation Results}
In the previous section, the central server collected interaction vectors from various clients and trained a recommendation model, resulting in recommendation vectors for each client/user. In this section, we will design an interactive protocol to send the recommendation results.

To protect each user's recommendation results from being intercepted by other users, we first use the vector splitting algorithm described in Section \ref{subsec: Splitting Interaction Vector} (Algorithm \ref{Alg:Splitting interaction vector}) to split the recommendation results into tuples $(\bm{V}_{Rec,i}^s, \bm{u}_{Rec,i,1}^*)$. These tuples, along with the virtual ID, form the triples $(V_{ID}^i, \bm{V}_{Rec,i}^s, \bm{u}_{Rec,i,1}^*),i=1,2, \ldots, N_{user}$.

The central server randomly sends these triples to different clients. Each client that receives the triple compares the $V_{ID}^i$ with the locally generated virtual ID. If they match, the client stores the triple locally. If they do not match, the client randomly forwards the triple to another client.

After multiple iterations, each client can receive all the triples belonging to itself (confirmed by comparing the virtual ID $V_{ID}^i$). Finally, each client will use the tuples to compute the interaction information, i.e.,
\begin{eqnarray}\label{u_Rec,i^*}
\mathbf{u}_{Rec,i}^* =
\left[\begin{array}{c}
    \bm{u}_{Rec,i,1}^* \\
    \sum _s \bm{V}_{Rec,i}^s
\end{array}\right].
\end{eqnarray}
Finally, remove the redundant information to obtain $\tilde{\bm{u}}_{Rec,i}^*$, that is, remove the columns containing $0$ from the matrix $\mathbf{u}_{Rec,i}^*$, then delete the second row of the matrix. Denote the resulting matrix as $\tilde{\bm{u}}_{Rec,i}^*$. Thus, $\tilde{\bm{u}}_{Rec,i}^*$ represents the recommended items of client/user $i$, equivalent to $\bm{Rec}_i$.

\begin{algorithm}[!t]
	\caption{Sending Recommendation Results}
	\label{Alg:Sending Recommendation Results}
	\begin{flushleft}
		\textbf{Input: } Each client corresponds to a recommended item $\bm{Rec}_i$, the central server receives triplets $(V_{ID}^i, \bm{V}_i^s, \bm{u}_{i,1}^*)$ in Algorithm \ref{Alg:Collecting Interaction Vectors}, $i=1,2, \ldots, N_{user}$.
	\end{flushleft}
	\begin{algorithmic}[1]
        \STATE \texttt{Central server do:}
        \STATE Split vectors $\bm{Rec}_i$ using Algorithm \ref{Alg:Splitting interaction vector} and form triples $(V_{ID}^i, \bm{V}_{Rec,i}^s, \bm{u}_{Rec,i,1}^*)$ with $V_{ID}^i$ obtained from Algorithm \ref{Alg:Collecting Interaction Vectors}. Then, send them to random clients individually.

        %\STATE \texttt{Receive and send data:}
        \WHILE{$1$}
            \STATE \texttt{Client $i$ do:}
            \STATE Receive the triplet $(V_{ID}^j, \bm{V}_{Rec,i}^s, \bm{u}_{Rec,i,1}^*)$.
            \IF{$V_{ID}^j == V_{ID}^i$}
                \STATE Save the triplet $(V_{ID}^j, \bm{V}_{Rec,i}^s, \bm{u}_{Rec,i,1}^*)$ in the local database $LD_{Rec}$,
            \ELSIF {$V_{ID}^j \in LD$}
                \STATE Send $(V_{ID}^j, \bm{V}_{Rec,i}^s, \bm{u}_{Rec,i,1}^*)$ to the client with the IP address $I_{IP}^j$,
            \ELSE
                \STATE Randomly send the triplet $(V_{ID}^j, \bm{V}_{Rec,i}^s, \bm{u}_{Rec,i,1}^*)$ to any client.
            \ENDIF
        \ENDWHILE
        %\definecolor{shadecolor}{rgb}{0.5,0.92,0.5}
%\begin{shaded}
        %\STATE \texttt{Aggregate recommended items}
        \STATE \texttt{Client $i$ do:}
        \STATE Use Eq.~\eqref{u_Rec,i^*} to calculate $\mathbf{u}_{Rec,i}^*$, and remove the redundant information to obtain $\tilde{\bm{u}}_{Rec,i}^*$.
%\end{shaded}

	\end{algorithmic}
        \textbf{Output: } The recommended items $\tilde{\bm{u}}_{Rec,i}^*$ for client/user $i$.
\end{algorithm}

In fact, in this method, randomly sending triples makes it difficult to deliver the triples to the correct client. To address this issue, during the collection of interaction vectors in Section \ref{subsec: Collecting Interaction Vectors}, we established the tuple $(V_{ID}^j, I_{IP}^j)$ to improve sending efficiency. When a client identifies that the $V_ {ID}^i$ in the received triple $(V_{ID}^i, \bm{V}_{Rec,i}^s, \bm{u}_{Rec,i,1}^*)$ matches the $V_ {ID}^i$ in a tuple $(V_{ID}^j, I_{IP}^j)$ in the local database, it indicates that the client with IP address $I_{IP}^j$ had previously sent a triple containing the virtual ID $V_ {ID}^i$ to the current client during the collecting interaction vectors in Section \ref{subsec: Collecting Interaction Vectors}. Therefore, the tuple $(V_{ID}^j, I_{IP}^j)$ should be prioritized for sending to that client (instead of being sent randomly). Based on this principle, the random sending can be optimized into a structured message transmission path, which is the inverse of the random message sending path in Section \ref{subsec: Collecting Interaction Vectors}. This ensures that the triple $(V_{ID}^i, \bm{V}_{Rec,i}^s, \bm{u}_{Rec,i,1}^*)$ is quickly sent to the correct client. The entire process of sending recommendation results is outlined in Algorithm \ref{Alg:Sending Recommendation Results}.

\section{Experiments}\label{sec: Experiments}
In this section, we will experimentally verify the security, transmission accuracy, and communication cost of the proposed privacy-preserving computation method.

\subsection{Experimental Settings}
In this experiment, we conducted evaluations based on the Yelp2018 dataset to assess the effectiveness and security of the proposed algorithm. The Yelp2018 dataset was used in the $2018$ challenge held by Yelp, containing users' rating information for various businesses, such as restaurants and bars. This dataset includes $31,831$ users, $40,841$ items, and $1,666,869$ interactions, with a density of $0.00128$~\cite{DBLP:conf/sigir/Wang0WFC19}. We will conduct the evaluation from the following perspectives.

\subsubsection{Security}
Regarding security, we test the proposed method's security by verifying whether obtaining some split vectors $\bm{V}_i^s$ can reconstruct the interaction vector $\bm{u}_i$. Specifically, we assume that several split vectors $\bm{V}_i^s$ of an interaction vector $\bm{u}_i$ are obtained by a certain client $i$, and we explore how many split vectors this client needs to approximately recover the interaction vector $\bm{u}_i$. When client $i$ obtains $t$ split vectors $\bm{V}_i^s$, it can use equation \eqref{u_i,2^**} to calculate the sum of $\bm{V}_i^s$ to obtain the speculated vector $\bm{u}_{spe}$, denoted as
\begin{eqnarray}\label{u_spe}
\bm{u}_{spe} = \sum _{s<t} \bm{V}_i^s.
\end{eqnarray}
By calculating the Jaccard similarity between the speculated vector $\bm{u}_{spe}$ and the interaction vector $\bm{u}_{i,2}^{**}$, we can determine the amount of interaction information obtained by client $i$. This experiment considers two scenarios. The first scenario examines the relationship between the number of split vectors $\bm{V}_i^s$ obtained by client $i$ and the Jaccard similarity when the total number of split vectors is $50$, $100$, and $200$. The second scenario examines the relationship between the ratio of fake interaction items to real interaction items, denoted as $c$, and the Jaccard similarity. The ratios considered are $c=2, 4, 6, 8, 10$.

\subsubsection{transmission accuracy}
Regarding transmission accuracy, the proposed privacy-preserving computation method needs to accurately send interaction information to the central server. During this process, if the randomly generated virtual IDs by the clients are duplicated, it will result in errors when the central server aggregates the interaction information. Therefore, to ensure high accuracy, the repetition rate of virtual IDs among all clients must be minimized. In this experiment, we will compare the relationship between the length of virtual IDs and the repetition rate.

\subsubsection{Communication cost}
For the efficiency and practicality of the interaction process, we conducted the following experiment with groups of $1000$ clients.

Regarding communication cost, we first considered the impact of the attenuation factor $\alpha$ on communication cost during the collecting interaction vectors phase and the sending recommendation results phase. Based on this, we selected the optimal value for the attenuation factor. Furthermore, we examined the relationship between the total communication cost in bytes and the number of clients, which is a crucial metric when considering communication costs in privacy-preserving computation.

Additionally, since the communication frequency of the proposed method is influenced by both probability and the number of clients, it is necessary to further verify whether the method is suitable for scaling to large data applications. We conducted experiments to compare the relationship between the number of clients participating in the computation during data interaction and the average number of information transmissions per client (i.e., the number of times triplets are sent).

\subsection{Results and Analysis}
\subsubsection{Results of the Safety Experiment}
\begin{figure*}[!t]%
\centering
\includegraphics[width=1.0\textwidth]{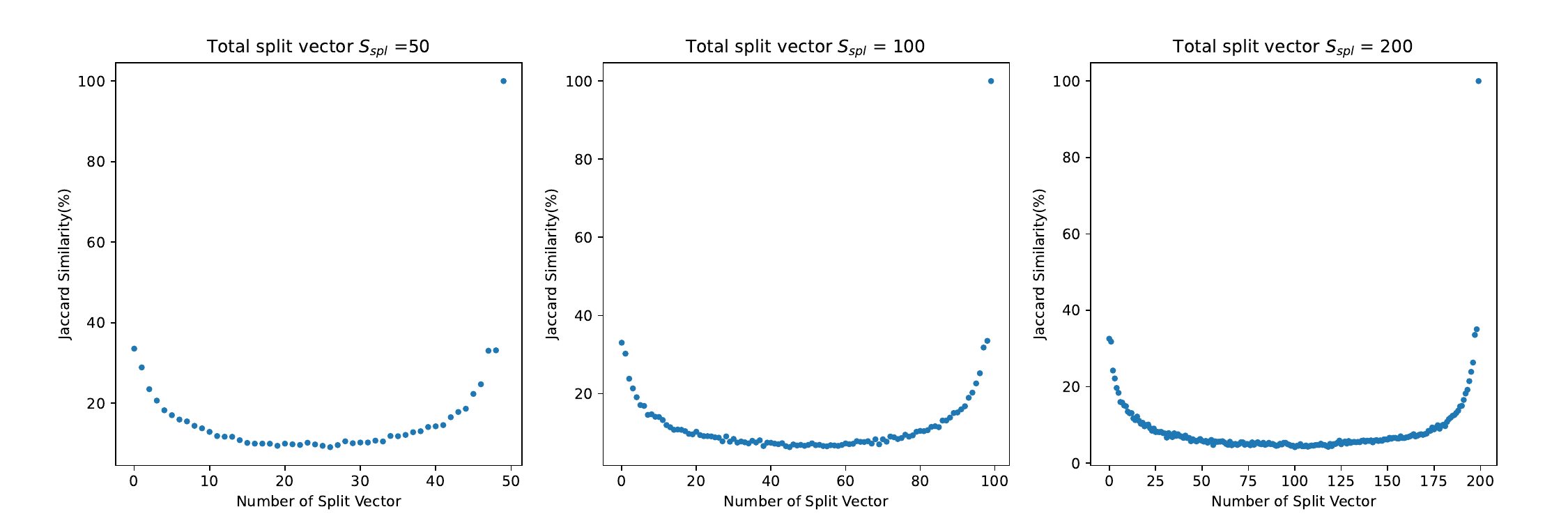}\\
\caption{The Jaccard similarity between the speculated vector $\bm{u}{spe}$, calculated by client $i$ using split vectors, and the interaction vector $\bm{u}{i,2}^{**}$.}\label{Fig-sim-num}
\end{figure*}

Figure~\ref{Fig-sim-num} shows the results of the first security experiment. The three subfigure show the experimental results for total split vectors of $50$, $100$, and $200$, respectively. As shown in \figurename~\ref{Fig-sim-num}, client $i$ can only achieve $100\%$ recovery of the interaction vector $\bm{u} _{i,2}^{**}$ (i.e., Jaccard similarity of $100\%$) if it receives all the split vectors. When the number of split vectors received by client $i$ is less than the total number, the Jaccard similarity is less than $40\%$. This experimental result indicates that any client attempting to steal another client's interaction information must obtain all the split vectors; otherwise, it cannot reconstruct the interaction vector $\bm{u}_{i,2}^{**}$, ensuring the security of the interaction information. Moreover, since each client is unaware of the number of split vectors corresponding to the interaction vectors of other clients, the true interaction information is further protected from being leaked.

\begin{figure*}[!t]%
\centering
\includegraphics[width=0.7\textwidth]{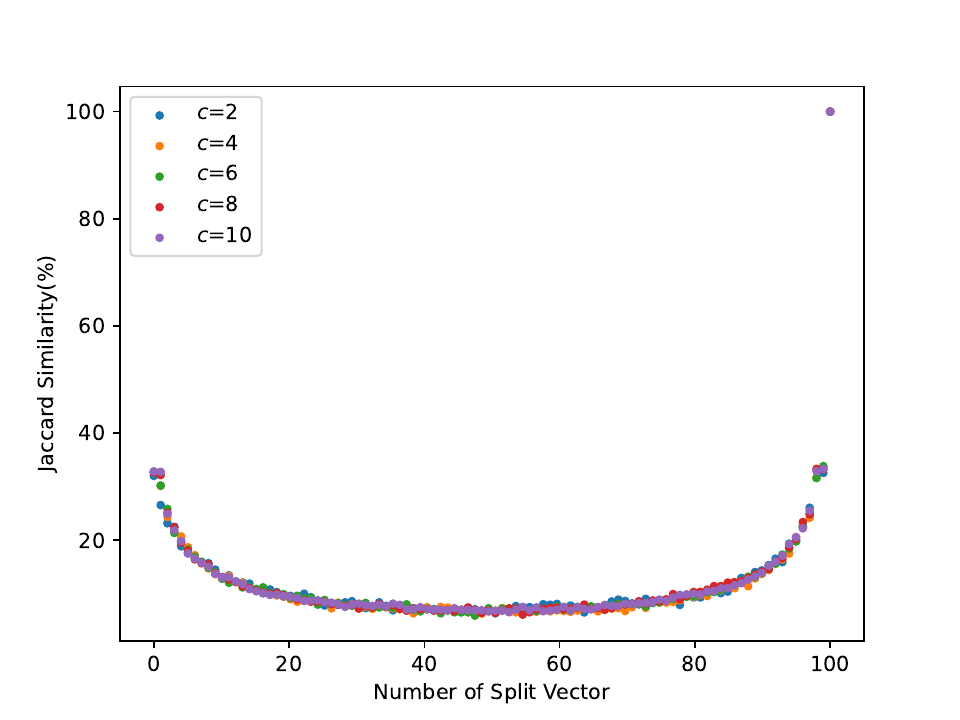}\\
\caption{The impact of the ratio $c$ of fake interaction items to real interaction items on the Jaccard similarity.}\label{Fig-sim-M}
\end{figure*}

Figure~\ref{Fig-sim-M} shows the results of the second security experiment. The ratio $c$ of fake interaction items to real interaction items affects the likelihood of real interaction items being leaked. Theoretically, the larger the value of $c$, the more fake interaction information is included in the vector $\bm{u}_{i,1}^*$, making it easier to hide the real interaction information. Figure~\ref{Fig-sim-M} shows the impact of different $c$ values on the Jaccard similarity. The experimental results indicate that setting different values of $c$ does not significantly affect the Jaccard similarity, client $i$ needs to collect all the split vectors to recover the interaction information. This experiment demonstrates that it is not necessary to set a large number of fake interaction items to effectively protect the real interaction items.

\subsubsection{Results of Transmission Accuracy Experiment}

\begin{figure*}[!t]%
\centering
\includegraphics[width=0.7\textwidth]{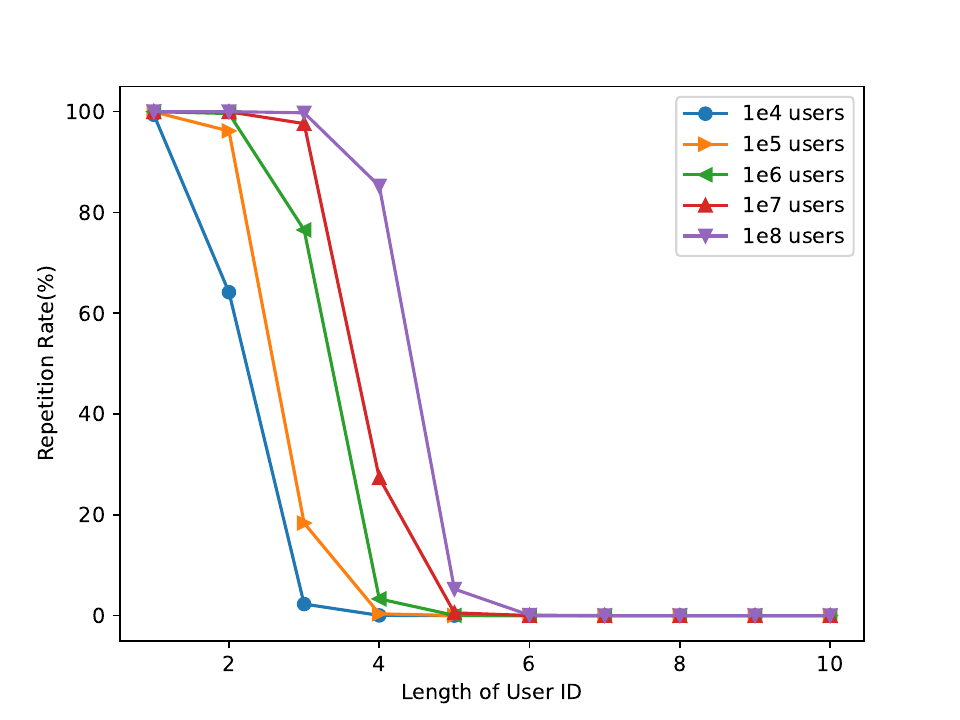}\\
\caption{The relationship between the number of digits in the virtual IDs and the repetition rate.}\label{Fig-IDLength-repeatRate}
\end{figure*}

Figure~\ref{Fig-IDLength-repeatRate} shows the results of the accuracy experiment. As shown in \figurename~\ref{Fig-IDLength-repeatRate}, increasing the number of digits in the virtual IDs can reduce the repetition rate, thereby preventing the central server from incorrectly aggregating the interaction vectors. Additionally, the experiment indicates that the repetition rate can be reduced to $0\%$ with virtual IDs of just $7$ characters, with ASCII codes ranging from $48$ to $57$ (digits), $65$ to $90$ (uppercase letters), and $97$ to $122$ (lowercase letters).

\subsubsection{Results of Communication Cost Experiment}
\begin{figure*}[!t]%
\centering
\includegraphics[width=0.7\textwidth]{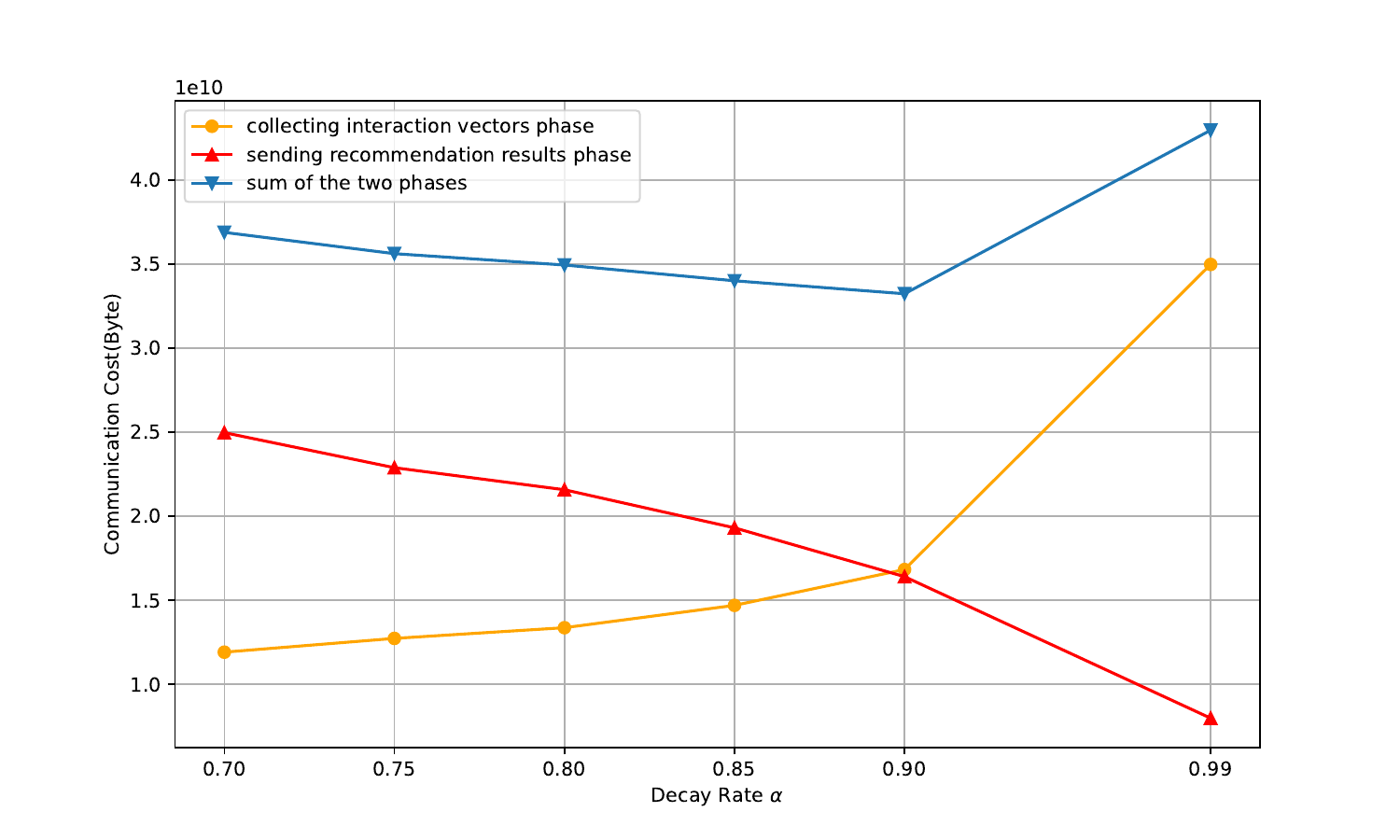}\\
\caption{The impact of the attenuation factor $\alpha$ on communication cost during the collecting interaction vectors phase and the sending recommendation results phase.}\label{Fig-communicationCost-decayRate}
\end{figure*}

Figure~\ref{Fig-communicationCost-decayRate} shows the results of the first communication cost experiment. From this figure, we can draw the following three conclusions:

\begin{itemize}
\item During the collecting interaction vectors phase, the communication cost increases with the attenuation factor $\alpha$. This is because a larger attenuation factor increases the probability of clients participating in communication (as indicated in lines $6$ to $12$ of Algorithm~\ref{Alg:Collecting Interaction Vectors}), thereby increasing the number of communications and the communication cost.

\item During the sending recommendation results phase, the communication cost decreases with the increase of the attenuation factor. This is because the tuples $(V_{ID}^j, I_{IP}^j)$ that we set indirectly reduce the communication cost. Specifically, with a larger attenuation factor, the number of communications during the collecting interaction vectors phase increases, thus increasing the probability that each client stores other clients' ID information tuples $(V_{ID}^j, I_{IP}^j)$ in the local database $LD$ (as indicated in line $18$ of Algorithm~\ref{Alg:Collecting Interaction Vectors}). Therefore, in the sending recommendation results phase, each client can send information based on the tuples $(V_{ID}^j, I_{IP}^j)$, increasing the probability of correctly sending information and reducing the communication cost.

\item From the sum of the communication costs of the two phases, it can be seen that the total communication cost is minimized when the attenuation factor $\alpha=0.90$.
\end{itemize}

\begin{figure*}[!t]%
\centering
\includegraphics[width=0.7\textwidth]{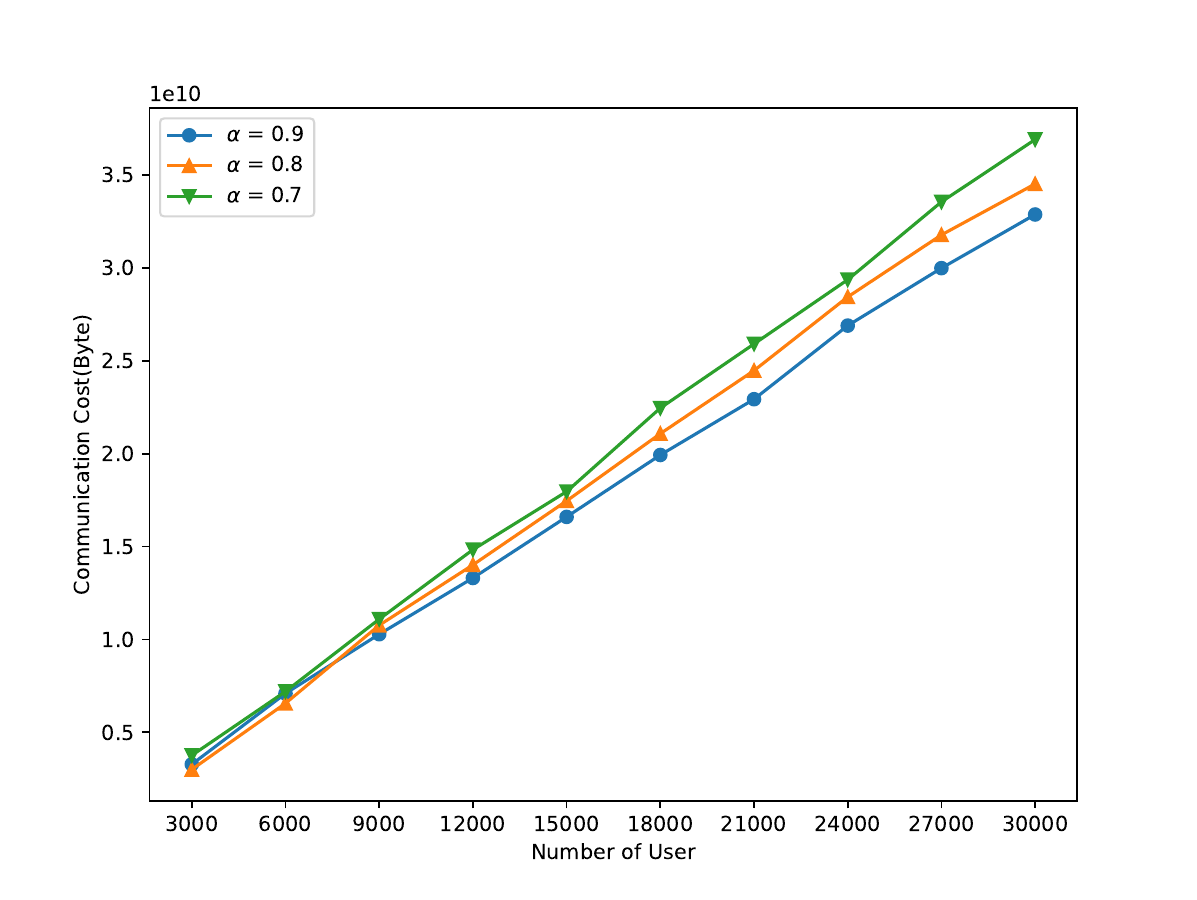}\\
\caption{The impact of the number of clients participating in privacy-preserving computation on the total communication cost.}\label{Fig-communicationCost-user}
\end{figure*}

Figure~\ref{Fig-communicationCost-user} shows the results of the second communication cost experiment. As shown in \figurename~\ref{Fig-communicationCost-user}, under different attenuation factor settings, the total communication cost of the proposed privacy-preserving computation method has a roughly linear relationship with the number of clients. This indicates that if this method is extended to large-scale data privacy computations, it will not lead to an exponential increase in communication cost.

\begin{figure*}[!t]%
\centering
\includegraphics[width=0.7\textwidth]{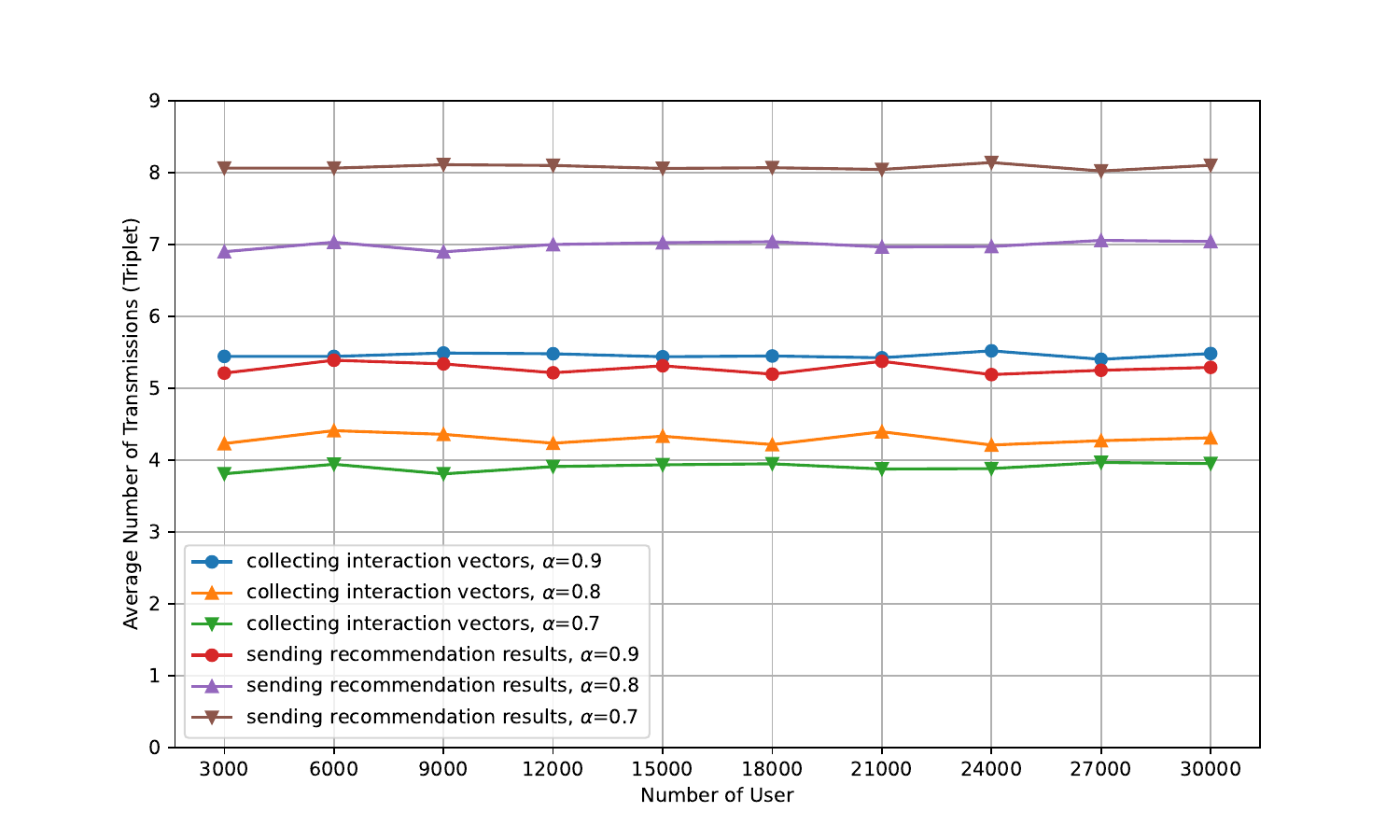}\\
\caption{The impact of the number of clients on the average number of transmissions (triplet).}\label{Fig-AverageTransmission-user}
\end{figure*}

Figure~\ref{Fig-AverageTransmission-user} shows the average number of triplets sent by each client during the collecting interaction vectors phase and the sending recommendation results phase. The experimental results indicate that as the number of clients increases, the average number of transmissions remains very stable, further validating the stability of the proposed method.

\section{Conclusion}
\label{sec: Conclusion}
%\cite{li_scaling_2014}~\cite{DBLP:journals/tnn/ChenSJ20, DBLP:journals/tnn/SattlerWMS20, DBLP:journals/tnn/XuDJHC22}
In this paper, we propose a new privacy-preserving recommendation system to address the security issues in recommendation systems and the scenario where each client only holds interaction data for a single user. We used split vectors and generated fake interaction items to protect the real interaction information of users and the recommendation results of the central server. During the data upload (central server collecting data) and data download (central server sending recommendation results) processes, we designed interactive protocols, triplets, virtual IDs, and real IP addresses, among other methods and concepts, to ensure the security of private information. Compared to existing FedRec methods, our approach has an additional advantage: it can be flexibly integrated with mainstream recommendation models without the need to design specific privacy-preserving computation methods tailored to the characteristics of each model. In the future, we plan to further explore the integration of this method with federated learning techniques, aiming to combine the strengths of both approaches to design a more practical and valuable privacy-preserving recommendation system.





%\begin{thebibliography}{99}
%\bibitem{}
\bibliographystyle{elsarticle-num}
\bibliography{PPRSref}
%\end{thebibliography}
\end{document}